\documentclass[preprint,12pt]{elsarticle}
\usepackage{amssymb}
\usepackage{comment}
\usepackage{multirow}
\usepackage{multicol}
\usepackage{adjustbox}
\usepackage{xcolor}
\usepackage{booktabs}
\usepackage{multirow}
\usepackage{comment}
\usepackage[colorlinks]{hyperref}
\usepackage{geometry}
\newgeometry{left=2cm,right=2cm,top=2cm,bottom=2cm}
\journal{Drug and Alcohol Dependence}
\begin{document}
\begin{frontmatter}

\title{The Impact of Opioid Prescribing Limits on Drug Usage in South Carolina: A Novel Geospatial and Time Series Data Analysis}

\author[inst1]{Amirreza Sahebi Fakhrabad}
\author[inst1]{Amir Hossein Sadeghi}
\author[inst2]{Eda Kemahlioglu-Ziya}
\author[inst2]{Robert B. Handfield}
\author[inst3]{Hossein Tohidi}
\author[inst3]{Iman Vasheghani Farahani}

\affiliation[inst1]{
            organization={Department of Industrial and Systems Engineering, North Carolina State University},
            addressline={915 Partners Way}, 
            city={Raleigh},
            postcode={27606}, 
            state={NC},
            country={USA}}
            
\affiliation[inst2]{ 
            organization = {Department of Business Management Poole College of Management, North Carolina State University},
            addressline={2801 Founders Dr}, 
            city={Raleigh},
            postcode={27607}, 
            state={NC},
            country={USA}}
            
\affiliation[inst3]{
            organization={SAS Institute Inc.},
            addressline={100 SAS Campus Dr},
            city={Cary},
            postcode={27513},
            state={NC},
            country={USA}}

\begin{abstract}
\paragraph{\textbf{Background}}
To curb the opioid epidemic, legislation regulating the amount of opioid prescriptions covered by Medicaid (Title XIX of the Social Security Act Medical Assistance Program) came into effect in May 2018 in South Carolina.

\paragraph{\textbf{Methods}}
We employ a classification system based on distance and disparity between dispensers, prescribers, and patients and conduct an ARIMA analysis on each class and without any class to examine the effect of the legislation on opioid prescriptions, considering secular trends and autocorrelation. The study also compares trends in benzodiazepine prescriptions as a control.

\paragraph{\textbf{Results}} 
The proposed classification system clusters each transaction into 16 groups based on the location of the stakeholders. These categories were found to have different prescription volume levels, with the highest group averaging 96.50 in daily MME (95\% CI [63.43, 99.57]) and the lowest 37.78 (95\% CI [37.38,38.18]). The ARIMA models show a decrease in overall prescription volume from 53.68 (95\% CI [53.33,54.02]) to 51.09 (95\% CI [50.74,51.44]) and varying impact across the different classes.

\paragraph{\textbf{Conclusion}} 
Policy was effective in reducing opioid prescription volume overall. However, the volume of prescriptions filled where the prescribing doctor is located more than 1000 miles away from the patient went up, hinting at the possibility of doctor shopping.

\end{abstract}

%%Research highlights
\begin{highlights}
\item To curb the opioid epidemic, states implemented opioid prescribing limits.
\item A novel geospatial model is proposed for classifying each opioid transaction.
\item The impact of opioid prescription limits varies among different proposed classes.
\item Geospatial model helps capture potential shortcomings of prescription limit policies.
\end{highlights}


\end{frontmatter}

\section{Introduction}
Illicit drug use, smoking, and alcohol in total kill 11.8 million people each year, which exceeds the number of deaths from all cancers worldwide and motor vehicle deaths in the United States \citep{roth_global_2018, Ciccarone2019}. Illicit drugs are drugs that have been prohibited under international drug control treaties and opioids are one of them \citep{babor_drug_2009}. In 2020, 91,799 drug overdose deaths occurred in the United States and the age-adjusted rate of overdose deaths increased by 31\% from 2019 (21.6 per 100,000) to 2020 (28.3 per 100,000), this happened while the Centers for Disease Control and Prevention (CDC) had declared that prescription drug abuse, including misuse of opioid medications, is at epidemic levels \citep{hedegaard_drug_2022,centers2011prescription}. The percentage change in the number of drug overdose deaths in South Carolina from July 2021 (1,959 cases) to July 2022 (2,173 cases) is 10.92\%, which makes South Carolina one of the states with high drug overdose deaths rate in the U.S. \citep{ahmad2021provisional}. While during the same period of time the national rate is 0.9\% \citep{ahmad2021provisional}.

Many states have implemented prescription limits on certain scheduled drugs, such as opioids, in an effort to reduce the quantity of these medications that are available and to address the issue of opioid abuse and overdose \citep{jackson2020characterizing, reid2019mandatory}. These laws vary by state, so the specific details of the prescription limits and which drugs are subject to the limits can differ depending on the location \citep{jackson2022state, day2019social}.

Legislation that imposes limits on the quantity of opioid prescriptions that can be issued by healthcare providers is relatively recent \citep{sedney2021assessing}. In 2018, South Carolina implemented the first drug limits on the quantity of opioid prescriptions that can be issued by healthcare providers. Under this law, the maximum supply of opioids that can be prescribed is generally limited to a five-day supply or a daily maximum of 90 morphine milligram equivalents (MMEs), unless the patient is receiving treatment for chronic pain, cancer pain, pain related to sickle cell disease, hospice care, palliative care, or medication-assisted treatment for substance use disorder \citep{louca2022spillover}.

There are researches showing that the implementation of opioid-prescribing policies indeed decreases the amount of opioid prescriptions as well as overdoses resulting from them \citep{BUONORA2022103888, beaudoin_state-level_2016}. However, other researchers have revealed that current state laws can unintentionally encourage people with opioid use disorders to seek the illicit market for drugs, thus increasing the overdose rate and suggesting that these types of legislation are not a strong deterrent \citep{lee_systematic_2021, BOWEN201990, Martins2019}. A systematic review of health information technology reveals that patients with drug misuse seek out pharmacies that do not participate in Prescription Drug Monitoring Programs (PDMPs). These patients travel long distances in order to engage in doctor shopping and obtain multiple prescriptions for controlled substances without detection, and paying in cash to do so \citep{kruse_health_2020, simeone_doctor_2017, cepeda_comparison_2013,cepeda_distance_2013}.

In this paper, we present an analysis of the characteristics of key stakeholders and state policies in South Carolina. To evaluate the effectiveness of opioid prescription limit policies, we propose a spatiotemporal classification system that takes into account the location of the three main players in the opioid supply chain (prescribers, dispensers, and patients) and the total distance traveled by patients. This system is used to identify groups with unusually high opioid prescription rates (measured by MME per day) and to evaluate the potential drawbacks of the state-level policy on different groups. Additionally, the overall impact of the policy on the amount of overall opioid prescriptions is evaluated, independent of the classification system. To the best of our knowledge, no one has analyzed the impact of opioid policies using clustering methods based on distance.

\section{Materials and methods}
\subsection{Research Design, Data Source}
This study developed a classification system based on the location of and distance traveled by stakeholders in prescription records to identify patterns and trends and detect potentially illicit activity in the opioid supply chain. We then used an interrupted time series quasi-experimental design to evaluate the impact of a prescription limit policy on opioid prescribing practices in South Carolina. This approach allowed us to examine the policy's effects on different subgroups within the population, such as the location of each player (patient, prescriber, and dispenser), and to understand the overall impact on opioid prescribing practices in the state. To control for temporal trends in prescribing of similar medications, we also examined the volume of incident benzodiazepine prescriptions as a controlled substance that is not expected to be affected by the opioid policies \citep{sedney2021assessing}.

\begin{figure}[h]
\centering
\includegraphics[scale=0.4]{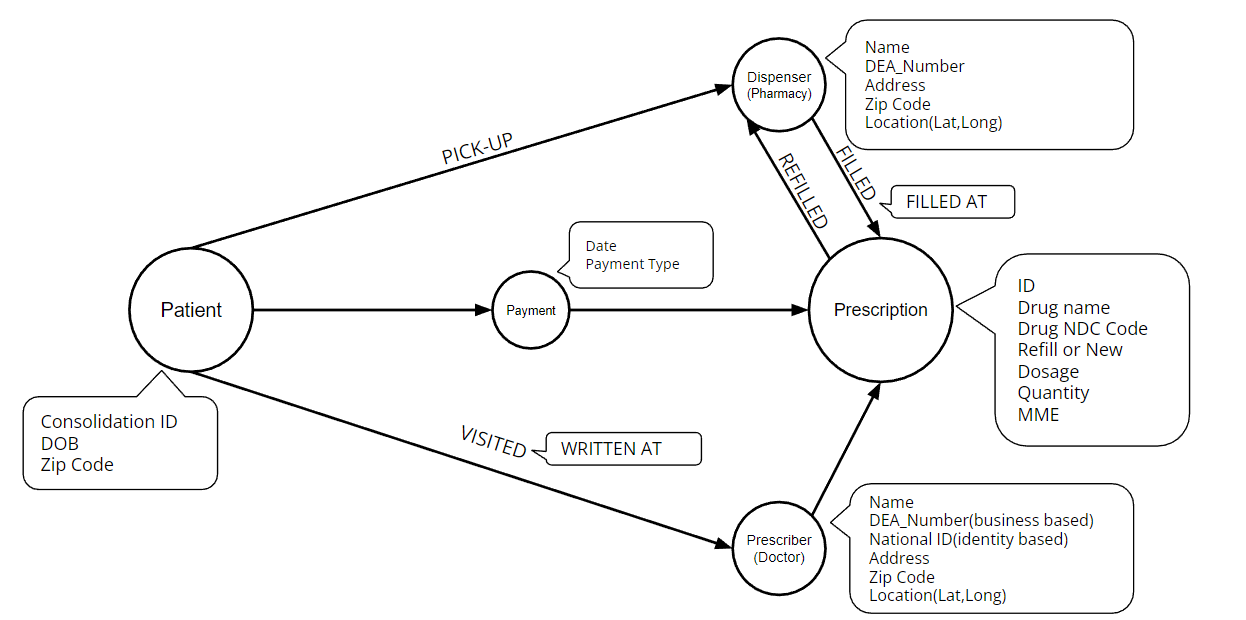}
\caption{Metagraph of the SCRIPTS dataset.}
\label{fig:Metagraph}
\end{figure}

In this study, data obtained from South Carolina Reporting and Identification Prescription Tracking System (SCRIPTS) is used \citep{noauthor_SCRIPTS_nodate}. The database contains 46 million prescription records from 2014 to 2022, including information about patients, prescribers, and dispensers for each particular opioid prescription. The interactions between them are analyzed using a network representation in Figure~\ref{fig:Metagraph}.

\subsection{Outcomes - Dependent Variable}
The primary focus of this study is the daily dosage of MME per prescription as a means of quantifying drug volume. MME is a unit of measurement that translates the total daily dosage of opioid medications into an equivalent dose of morphine for the purpose of comparing the relative strength of different opioids. This calculation is performed using a standard formula which converts the daily dose of each opioid into a morphine equivalent and adds these amounts together to determine the total MME. This allows healthcare providers to evaluate the potential risk of overdose or other adverse effects associated with opioid use. As recommended by CDC, prescriptions with daily MME values exceeding 50 MME require careful assessment, while those exceeding 90 MME are generally considered to be high risk and should be avoided \citep{dowell2022cdc, dowell2016cdc}. In order to express MME in daily terms, the total MME is divided by the number of days that the prescription is intended to last, as indicated by the "days supply" field in the SCRIPTS data set. Prescriptions with missing days supply data were excluded from analysis in this study.

\subsection{Exposures - Independent Variables}
\textit{NarxCare}: South Carolina has implemented several policies to reduce opioid misuse and abuse. One of these policies, called NarxCare, was implemented in May 2018 and limits opioid prescriptions to a five-day supply or 90 morphine milligram equivalents daily, with exceptions for severe illnesses. \citep{donnelly_south_2021, horwitz_problem_2018, noauthor_public_nodate} We will focus on studying the impact of this policy because it is the most recent and directly affects prescription dosage. In order to implement the policy impact in the model, we use a binary indicator variable (0 = pre-implementation, 1 = post-implementation).

\textit{Internal control series}: A dataset of benzodiazepine prescriptions was used as a control for comparison in the study because there is similar pressure to decrease benzodiazepine prescriptions but they were not specifically addressed in the NarxCare policy. All benzodiazepines dispensed from pharmacies in South Carolina were included in the dataset, including alprazolam, chlordiazepoxide, and others. These were used as a control because benzodiazepines are not specifically addressed in the NarxCare policy, despite the CDC recommending avoiding their use with opioid medications and in addition, they have a lower potential for addiction, which makes them a suitable candidate for comparison. \citep{jeffery2019rates}.

\subsection{Statistical Analyses}
\subsubsection{Classification System} \label{sec:Classification System}
The study employed a classification system that grouped prescription records based on the location of stakeholders (patients, prescribers, and dispensers) and the total distance traveled (indicated as $\pi$) to fill the prescription. The system was divided into four distance groups: less than 250 miles, between 250 and 500 miles, between 500 and 1000 miles, and over 1000 miles, and three disparity groups: patient isolated, prescriber isolated, dispenser isolated. The logic behind the classification system is that the average distance from the center of South Carolina to its border is about 250 miles, which means that transactions with a total distance of fewer than 500 miles (round-trip) fall within the state's border. Transactions with distances between 500 and 1000 miles fall within the borders of neighboring states, and those with distances more than 1000 miles are considered to be in non-neighboring states. The second logic is that typically, two or three out of the three stakeholders are close to each other. For example, it is normal to see a record in which a patient travels to visit a doctor, gets a prescription, and fills it in a pharmacy near the doctor or on the way home while visiting doctors or pharmacies located further away is not a common pattern. We used this classification system to identify patterns and trends in the supply chain of opioid prescriptions. Further details on how the disparity and distance level representations of the SCRIPTS network graph were developed can be found in the supplementary file.

\begin{table}[h]
\caption{Classification system illustration}
\vspace{0.2cm}
\label{tab:coding system definition}
\begin{adjustbox}{width=1.0\textwidth, center}
    \begin{tabular}{|clc|clc|}
    \hline
        \textbf{Distance (miles)} & \textbf{Disparity} &  \textbf{Code} & \textbf{Distance (miles)} & \textbf{Disparity} &  \textbf{Code}\\ 
         \hline \hline
         \multirow{4}{*}{$\pi \leq 250$}  &  Patient isolated & \textbf{00} & \multirow{4}{*}{$500 < \pi \leq 1000$} &  Patient isolated & \textbf{20} \\  
          &  Prescriber isolated & \textbf{01} & & Prescriber isolated & \textbf{21}\\ 
          &  Dispenser isolated & \textbf{02} &  & Dispenser isolated & \textbf{22}\\  
          &  Otherwise & \textbf{03} & & Otherwise & \textbf{23}\\ 
          \hline
         \multirow{4}{*}{$250 < \pi \leq 500$}  &  Patient isolated & \textbf{10} & \multirow{4}{*}{$\pi > 1000$} &  Patient isolated & \textbf{30} \\  
          &  Prescriber isolated & \textbf{11} & & Prescriber isolated & \textbf{31}\\ 
          &  Dispenser isolated & \textbf{12} &  & Dispenser isolated & \textbf{32}\\ 
          &  Otherwise & \textbf{13} & & Otherwise & \textbf{33}\\ 
          \hline
\end{tabular}
\end{adjustbox}
\end{table}

\subsubsection{Interrupted Time Series Analyses}
In the second step, to examine the impact of the policy on opioid prescribing practices of the proposed classes, we used an autoregressive integrated moving average (ARIMA) interrupted time series analysis (ITS). This statistical technique is particularly useful for evaluating the effectiveness of interventions, and the ARIMA model is a commonly used method in healthcare research. The ARIMA model is a statistical method used for time series analysis that was first introduced in 1976. It combines elements of both autoregressive (AR) and moving average (MA) models to forecast stationary and non-stationary data. In an AR model, the predicted variable is determined by its own previous values and an error term, while in an MA model, the predicted variable depends on the current and past values of random shock terms. The ARIMA model combines these two models and has the following form:

\begin{equation}
    y^{\prime}_t = c + (\phi_1 y^{\prime}_{t-1} + ... + \phi_p y^{\prime}_{t-p}) + (\theta_1 \epsilon_{t-1} + ... + \theta_q \epsilon_{t-q}) + \epsilon_t \\
\end{equation}

Where $c$ is a constant, $Y_i$ is time series value at time $i$, $\phi_1$, $\phi_2$, ..., $\phi_p$ are parameters of the AR model with p lags, $\epsilon_t$ is
normal random noise at time $t$, $\theta_1$, $\theta_2$, ..., $\theta_q$ are coefficients of the MA model with q orders, and $y^{\prime}_t = \Delta^d y_t$ is the $d^{th}$ order difference of $y_t$ helps to produce a stationary process. \\

The following steps were taken in the study:
\begin{enumerate}
    \item An ARIMA model was fitted for each time series under study, specifically for the pre-intervention period. The pre-intervention period is considered the period before the policy enactment in May 2018. 

    \item An autoregressive integrated moving average with exogenous variables (ARIMAX) model was estimated for the entire time series, adding regressor variables corresponding to the announcement and implementation of the legislation. This model was used to identify the intervention effect of the announcement and enactment of the law.

    \item All previous steps were repeated for both subgroups proposed by the classification system and for the aggregated data without any subgroup.

\end{enumerate}

This approach has been previously reported in the literature \citep{sedney2021assessing, branham2018time,  delcher2015abrupt, maierhofer2022intended, nguyen2022mandatory}. The Automatic Time Series Modeling (ATSM) package of SAS Visual Forecasting software (Viya 4 - Cadence version 2022.12) was used to find the best ARIMA models for each series \citep{fouda2020sas}. The software automatically detects the required differencing orders by running simple and seasonal augmented Dickey-Fuller tests. Then, it finds the tentative autoregressive moving average (ARMA) orders using the minimum information criterion. Next, the optimizer finds the best ARMA orders, bounded by their respective tentative values, considering the Bayesian information criterion (BIC). The seasonal ARMA orders are similarly detected while they are less than or equal to 2. The intervention effect on the response variable was analyzed by adding event inputs in shape of level-shift, ramp and inverse trend to the final ARIMA model for each series through the interactive modeling node of SAS Visual Forecasting \citep{version2000statistical}.

\section{Results}
\subsection{Overall observation of opioid prescriptions}
In this section, we demonstrate the efficiency of our proposed classification model in section~\ref{sec:Classification System} using two hypothesis tests. We test if the consumption pattern is significantly correlated with the distance traveled by patients and the relative location of other players.

We conduct a cross-sectional analysis of prescription records from the state of South Carolina. To ensure the validity of our findings, data prior to 2014 was excluded from the analysis due to high levels of noise. Additionally, prescriptions with unusual MME exceeding $10^5$ were removed from the dataset. The data was further consolidated at the monthly level to enable time series analysis in the following section. This was done because any other level of consolidation would result in intermittent and non-stationary data, which is not compatible with ARIMA models. Table~\ref{tab:classification results} provides a summary of key statistics, including the days of supply, MME, MME per day (calculated as MME divided by days of supply), and the class percentage (represented as the total number of records within a specific group, divided by the total number of records in the dataset) for opioid drugs.

\begin{table}[h]
\caption{Overview of key statistics in different coding system classes (opioid drugs only) after policy implementation}
\label{tab:classification results}
\vspace{0.2cm}
\begin{adjustbox}{width=1\textwidth, center}
    \begin{tabular}{cccccc}
    \hline 
Group Code & Mean (SD) Days Supply & Mean (SD) MME     & MME per day (95\% CI) & \begin{tabular}[c]{@{}l@{}}Percentage\\  of MME\end{tabular} \\ \hline \hline
00 \textbf{(N=2,890)}         & 14.89 (4.87)            & 802.32 (310.46)  & \textbf{48.88 (48.61, 49.15)}    
    & 12.05                                                        \\
01 \textbf{(N=2,890)}          & 15.05 (5.09)            & 771.95 (292.38)  & \textbf{46.90 (46.70, 47.10)}
    & 35.76                                                        \\
02 \textbf{(N=2,890)}         & 16.04 (4.91)            & 809.49 (304.97)  & \textbf{48.66 (48.20, 49.12)}
    & 3.87                                                         \\
03 \textbf{(N=2,890)}        & 15.4 (5.15)            & 800.95 (302.02)  & \textbf{47.96 (47.75, 48.16)}
     & 40.65                                                        \\ \hline
10 \textbf{(N=2,883)}        & 8.84 (3.74)            & 478.36 (362.17)  & \textbf{44.00 (43.43, 44.57)}
     & 0.19                                                         \\
11 \textbf{(N=2,889)}        & 14.89 (5.95)            & 815.39 (395.73)  & \textbf{47.51 (47.15, 47.87)}
     & 1.29                                                         \\
12 \textbf{(N=2,438)}        & 19.37 (7.62)            & 1064.53 (994.58)  & \textbf{55.26 (53.66, 56.86)}
    & 0.08                                                         \\
13 \textbf{(N=2,889
)}        & 14.06 (5.20)            & 770.80 (357.95)  & \textbf{49.08 (48.68, 49.47)}
    & 1.66                                                         \\ \hline
20 \textbf{(N=2,884)}        & 7.56 (3.38)             & 309.30 (219.66)   &\textbf{37.84 (37.44, 38.24)}
     & 0.16                                                         \\
21 \textbf{(N=2,868)}        & 13.42 (6.25)            & 704.96 (411.89)  & \textbf{46.40 (45.93, 46.93)}
     & 0.80                                                         \\
22 \textbf{(N=2,335)}        & 18.93 (6.97)            & 1110.24 (746.99)  & \textbf{91.71 (87.20, 96.22)}
    & 0.07                                                          \\
23 \textbf{(N=2,889)}        & 12.69 (5.09)            & 721.36 (384.68)  & \textbf{52.46 (51.91, 53.01)}
     & 1.17                                                         \\ \hline
30 \textbf{(N=2,888)}        & 7.66 (3.15)             & 322.93 (230.59)   & \textbf{37.78 (37.38, 38.18)}
     & 0.19                                                         \\
31 \textbf{(N=2,883)}        & 12.69 (5.03)            & 723.15 (1336.44)  & \textbf{47.50 (46.93, 48.07)}
     & 1.22                                                         \\
32 \textbf{(N=2,460)}        & 20.06 (7.51)            & 1374.09 (1153.52) & \textbf{96.50 (93.43, 99.57)}
    & 0.14                                                         \\
33 \textbf{(N=2,889)}        & 12.81 (4.67)            & 761.86 (375.38)  & \textbf{56.00 (55.46, 56.54)}
     & 1.87                                                         \\ \hline
\end{tabular}
\end{adjustbox}
\end{table}

The first observation is that most of the records (more than 90\% of total transactions) have less than 250 miles of total traveled distance (class “0X”). It means most of the transactions are not suspicious in terms of distance. Any other distance categories (“1X”, “2X”, and “3X”) have an almost equal share of the data (~3\% of the transactions). 

In terms of MME/day, as stated by CDC the Hazard Ratio (HR) for 1-19 MME/day (level 1), 20-49 MME/day (level 2), 50-99 MME/day (level 3) and any number above 100 MME/day (level 4) defines as 1, 1.44, 3.73, and 8.87 respectively \citep{dowell2022cdc}. Overall, according to CDC guidelines, any prescription with more than 90 MME/day should be avoided while the ones with more than 50 MME/day are required to be carefully assessed. In our data set, MME/day average is 46.85 with a standard deviation of 9.29 which falls in the second and third CDC classes and is over 50 for many records. Among different defined classes, five of them (classes 12, 22, 23, 32, and 33) have MME/day of more than 50 on average (Table~\ref{tab:classification results}). The higher MMEs are mostly in groups where patients travel more than 500 miles (probably to nearby states) to get their drugs from long-distance pharmacies which increases the probability of illicit activities. We state these observations as hypotheses 1 and 2:

$H_1$: \textit{The total distance traveled by the patient and the relative location of the prescribers and the dispensers is correlated to the patient's daily MME.}

$H_2$: \textit{Patients who travel more than 250 miles to obtain their opioid prescription from a distant pharmacy have a significantly higher usage rate compared to other patients.}

The first hypothesis tests the suitability of the introduced classification system. If we do not observe any differences between classes, then the proposed system is not meaningful anymore. To evaluate its effectiveness, we utilize the one-way ANOVA test\citep{mcdonald_handbook_2009}. One-way ANOVA is not sensitive to deviations from normality and works well in the case of a large sample size \citep{delacre2019taking}. As a result, ANOVA is a valid approach in this context and there is no need to perform other tests like Kruskal-Wallis (more statistical result are provided in supplementary material).

While there is no definitive evidence as to the source of diverted or misused opioids, the study of diversion of medically prescribed opioids is most robust in the acute opioid prescription phase \citep{bicket2017prescription}. In this regard, the second hypothesis aims to identify classes with a consumption level of more than 50 MME per day as a potential disorder of consumption behavior. According to guidelines from the CDC, any dosage above 50 MME per day is considered risky and should be carefully evaluated \citep{dowell2022cdc}. A one-sided t-test with 95\% confidence intervals was used, with the alternative hypothesis being MME per day above 50 or 90, and the null hypothesis being MME per day less than 50 or 90. Table~\ref{tab:classification results} shows that groups 12, 22, 23, 32, and 33 have 95\% confidence intervals above 50 MME per day, while all other classes are less than 50 MME per day. Additionally, class 32 has MME per day above 90, which is highly risky. These results validate the second hypothesis.

\subsection{Policy analysis using time series models}

The State of South Carolina implemented a policy called NarxCare on May 1, 2018, with the aim of addressing the opioid epidemic by limiting the number of opioid prescriptions that can be prescribed to patients \citep{donnelly_south_2021, noauthor_public_nodate}. The policy's impact was evaluated by analyzing data from SCRIPTS and comparing the volume of opioid prescriptions before and after the policy's implementation. A simple pre-post policy comparison shows that on average, the MME per day was reduced by 2.59 (from 53.68 (95\% CI, [53.33,54.02]) to 51.09 (95\% CI [50.74,51.44])) and the majority of groups in the proposed classification system experienced a significant reduction in the volume of prescribed opioids, except for groups 22 and 31 (as shown in Table \ref{tab:MMEperDay_PrePostPolicy}). To study this observation rigorously, a combination of ARIMA and ARIMAX models, as a class of interrupted time series models, were used to further analyze the policy's impact over time and how it has affected different groups.

\begin{table}[h]
\caption{The MME per day in each group after and before policy implementation.}
\label{tab:MMEperDay_PrePostPolicy}
\vspace{0.2cm}
\begin{adjustbox}{width=1\textwidth, center}
\begin{tabular}{clcclcclcclcc}
\toprule
\multicolumn{1}{l}{} &   & \multicolumn{11}{c}{\textbf{Distance Level}} \\ \cline{3-13} 

\multicolumn{1}{l}{} &   & \multicolumn{2}{c}{\textbf{Code 0X: $\pi \leq 250$}} &  & \multicolumn{2}{c}{\textbf{Code 1X: $250 < \pi \leq 500$}} &  & \multicolumn{2}{c}{\textbf{Code 2X: $500 < \pi \leq 1000$}} &  & \multicolumn{2}{c}{\textbf{Code 3X: $\pi > 1000$}}  \\ \cline{3-4} \cline{6-7} \cline{9-10} \cline{12-13}

\multicolumn{1}{l}{} &   & \multicolumn{2}{c}{\textbf{Mean}} &  & \multicolumn{2}{c}{\textbf{Mean}} &  & \multicolumn{2}{c}{\textbf{Mean}} &  & \multicolumn{2}{c}{\textbf{Mean}}  \\ 

\multicolumn{1}{l}{} &   & \multicolumn{2}{c}{\textbf{(95\% CI)}} &  & \multicolumn{2}{c}{\textbf{(95\% CI)}} &  & \multicolumn{2}{c}{\textbf{(95\% CI)}} &  & \multicolumn{2}{c}{\textbf{(95\% CI)}}  \\ \cline{3-4} \cline{6-7} \cline{9-10} \cline{12-13}

\multicolumn{1}{c}{}  &   & \multicolumn{1}{c}{\textbf{Pre-Policy}} & \multicolumn{1}{c}{Post-Policy} &  & \multicolumn{1}{c}{\textbf{Pre-Policy}} & \multicolumn{1}{c}{\textbf{Post-Policy}} &  & \multicolumn{1}{c}{\textbf{Pre-Policy}} & \multicolumn{1}{c}{\textbf{Post-Policy}} &   & \multicolumn{1}{c}{\textbf{Pre-Policy}} & \multicolumn{1}{c}{\textbf{Post-Policy}} \\ \midrule

\multirow{12}{*}{\rotatebox[origin=c]{90}{\textbf{Disparity Level}}} & \multirow{2}{*}{\textbf{Code X0: Patient isolated}} & 50.57  & 46.84  &  & 47.38 & 39.93  &  & 40.41  & 34.74 &  & 40.01 & 35.10 \\ 
&  & (50.25, 50.89) & (46.42, 47.26) &  & (46.58, 48.18) & (39.18, 40.68)    &  &  (39.95, 40.87) & (34.11, 35.37)&  & (39.45, 40.57) & (34.55, 35.65)     \\ 
\\
 & \multirow{2}{*}{\textbf{Code X1: Prescriber isolated}}  & 49.18 & 44.15  &  & 48.78 & 45.98  &  & 48.56  & 43.81  &  & 46.69   & 48.49 \\
 & & (48.91, 49.45) & (43.91, 44.39)  &  & (48.38, 49.18) & (45.33, 46.63)  &  & (48.03, 49.09)  & (42.91, 44.71) &  & (46.35, 47.03)  & (47.31, 49.67)  \\
 \\
 &  \multirow{2}{*}{\textbf{Code X2: Dispenser isolated}} & 49.14  & 48.08  &  & 58.43 & 51.92 &  & 85.07 & 99.76  &  & 97.10 & 95.78 \\        
 & & (48.69, 49.59)  & (47.21, 48.95)  &  & (55.87, 60.99) & (50.06, 53.78) &  & (80.84, 89.30) &  (91.22, 108.30)  &  &  (93.36, 100.84)  & (90.70, 100.86) \\  
 \\
 &  \multirow{2}{*}{\textbf{Code X3: Otherwise}} & 49.57 & 46.01  &  & 50.78 & 47.02 &  & 53.77 & 50.89  &  & 56.81 & 55.02 \\
& & (49.30, 49.84) & (45.73, 46.29)  &  & (50.35, 51.21) &  (46.34, 47.70)   &  & (53.16, 54.38) & (49.93, 51.85) &  & (56.23, 57.39)  & (54.05, 55.99) \\
\bottomrule
\end{tabular}
\end{adjustbox}
\end{table}

As a first step, an ARIMA model was applied to the pre-intervention data and used to predict the post-intervention prescription volume. The forecast results of the model, along with the 95\% prediction confidence interval, are illustrated in Figure \ref{fig:TS__Trend__NoClass} (a detailed explanation of the models and parameters can be found in the supplementary materials). The discrepancy between the predicted values and actual ones in the opioid data is an indication of the intervention's impact. In the absence of any intervention effect, it would be expected that the pre-policy and post-policy data would follow the same pattern. Notably, the discrepancy is not present in the benzodiazepines, which serve as the control group.

\begin{figure}[!h]
\centering
\includegraphics[scale = 0.7]{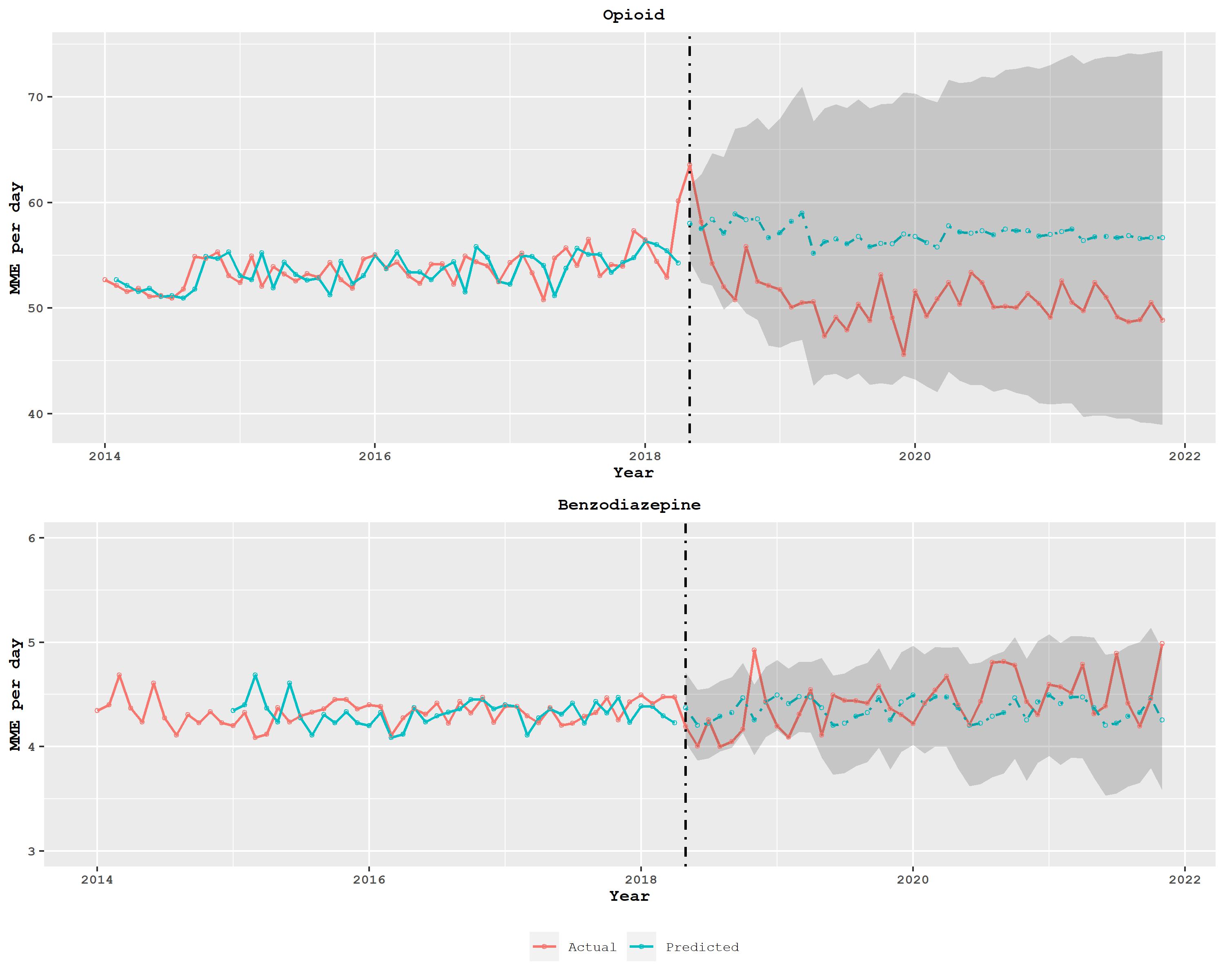}
\caption{The top graph shows the average daily MME for opioid prescriptions in the state of South Carolina over a period of time (measured in months). The broken vertical line indicates the legislative intervention that was enacted in May 2018. The blue dotted line in the graph represents the fit of the mathematical model. The bottom graph demonstrates the same for benzodiazepine as the control group.}
\label{fig:TS__Trend__NoClass}
\end{figure}

\begin{figure}[!h]
\centering
\includegraphics[scale = 0.6]{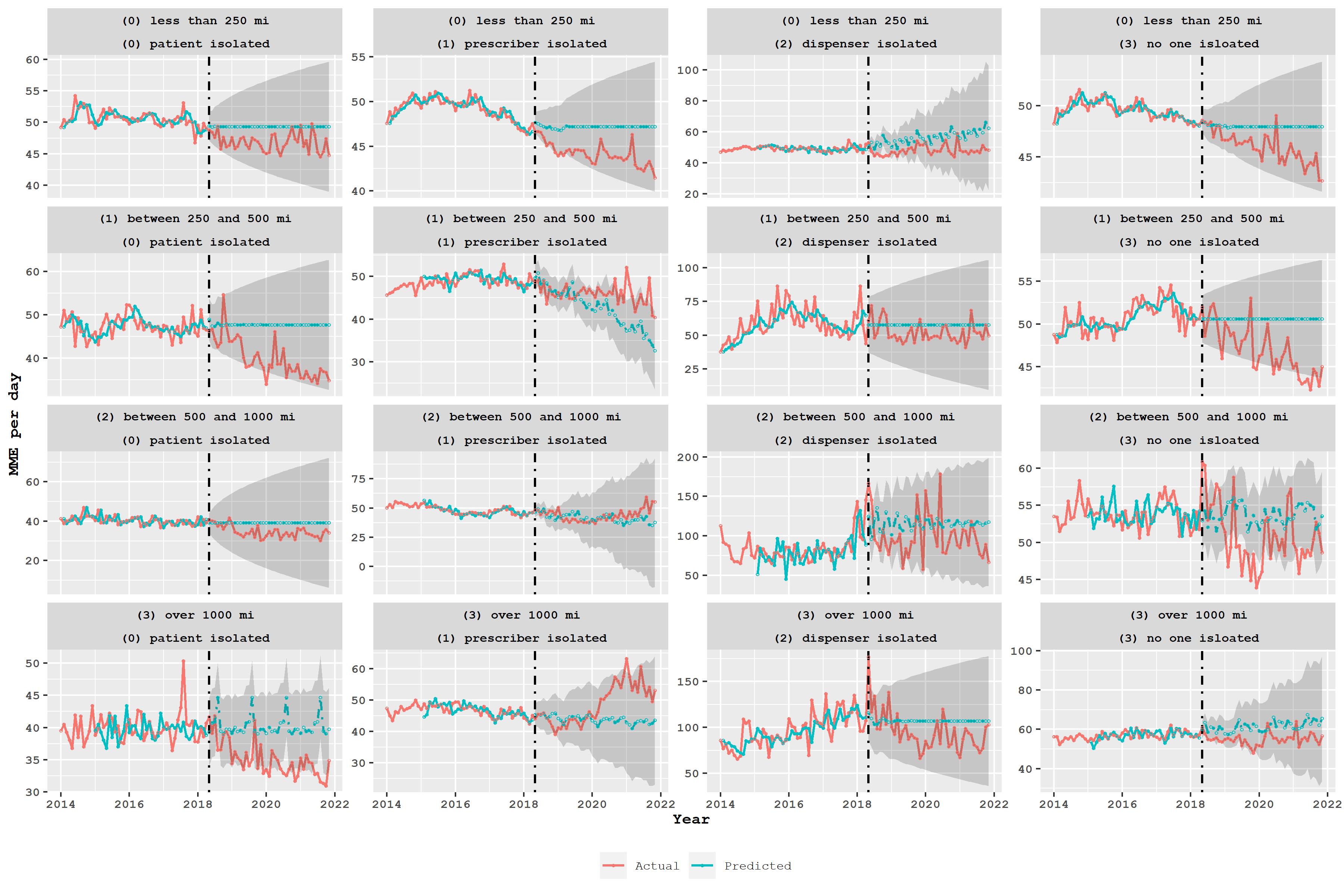}
\caption{Average daily MME for opioid prescriptions in the state of South Carolina over a period of time (measured in months) for different proposed classes. The broken vertical line indicates the legislative intervention that was enacted in May 2018. The blue line represents the fit of the mathematical model.}
\label{fig:TS__Trend__Opioid__Class}
\end{figure}

\begin{figure}[!h]
\centering
\includegraphics[scale = 0.6]{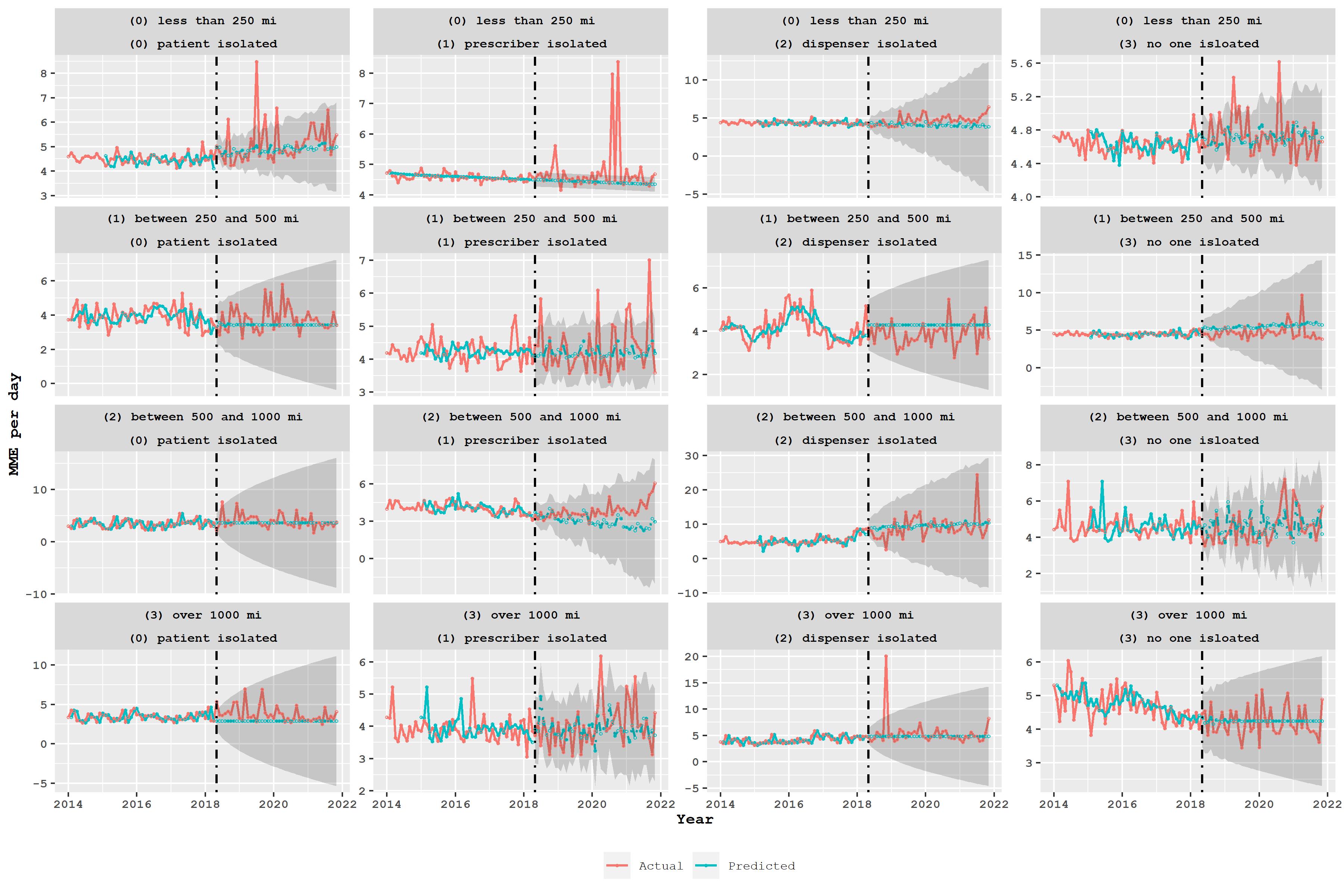}
\caption{Average daily MME for benzodiazepine (control group) prescriptions in the state of South Carolina over a period of time (measured in months) for different proposed classes. The broken vertical line indicates the legislative intervention that was enacted in May 2018. The blue line represents the fit of the mathematical model.}
\label{fig:TS__Trend__Benzos__Class}
\end{figure}

The same analysis was conducted at the class level to evaluate the impact of the policy on the groups of the proposed classification system, as shown in Figure \ref{fig:TS__Trend__Opioid__Class}. The desired mismatch, which is a reduction in actual values but no reduction in predicted values, is apparent for transactions with a total traveled distance less than 250 miles (first row) and patient-isolated classes (first column). However, when examining the prescriber-isolated classes (second column), it seems that the mismatch is occurring in the undesired direction, specifically in classes 11 and 31. This could be an indication of unintended consequences of the policy. The dispenser-isolated classes (third column) tend to have a higher MME per day, but there is no indication of reduction in post-policy predicted or actual values. In the last column, we can see the reduction in actual values and mismatch in groups 03 and 13. A similar analysis was also conducted for benzodiazepine prescriptions, and the results are presented in Figure \ref{fig:TS__Trend__Benzos__Class}. The policy does not appear to have a significant impact on the volume of benzodiazepine prescriptions, and the mismatch between predicted and actual values does not show any specific trend in many groups. (The details of the pre-policy fitted ARIMA models for each group are provided in the supplementary materials)

To validate the observations and evaluate the policy's impact mathematically, an ARIMAX model was used, considering the intervention as event inputs in the form of level-shift, ramp, and inverse trend to the previously fitted ARIMA models. As shown in Table \ref{tab:Overall Policy Effect}, it was found that the intervention has a significant impact on reducing opioid consumption in groups 03, 30, 11, 31, 13, and 23, while there is no impact on other opioid classes or benzodiazepines. These are important results since the desired impact in classes 03 and 30, which make up 40\% of the total transactions, may lead to the desired overall impact for opioid drugs. Moreover, there is still no desired policy effect over classes 22 and 32 with the highest volumes of prescriptions. Furthermore, there is an unintended impact of the policy over dispenser-isolated groups (class 11 and 31).

\begin{table}[!h]
\centering
\begin{adjustbox}{width=1\textwidth}
\tiny
\begin{tabular}{ccllllc}
\hline
\hline
\textit{\textbf{Class}}  & \textit{\textbf{Model}} & \textit{\textbf{Coefficient}} & \textit{\textbf{Estimate}} & \textit{\textbf{Standard Error}} & \textit{\textbf{P value}} & \textit{\textbf{Significance Level}}\\
\hline

\multirow{2}{*}{\textit{\textbf{Overall}}} & \multirow{2}{*}{ARIMA(1,1,0)} & \textit{AR (12)}   & $0.45$    & $0.10$      & $p<0.001$  & (***)  \\
 && \textit{Inverse Trend D(1)}   & $3241.52 $& $1098.68$    & $0.004$ & (**)\\
\hline

 \multirow{5}{*}{\textit{\textbf{11}}} & \multirow{5}{*}{ARIMA(1,1,1)} & \textit{Constant}   & $-5.28$    & $0.60$      & $p<0.001$ & (***)    \\
 && \textit{AR (12)}   & $-0.29$& $0.14$    & $0.040 $   &  (*)         \\
 && \textit{MA (1)}   & $1.00$& $0.03$    & $p<0.001$ & (***)            \\
 && \textit{Ramp}   & $-13.56$& $4.48$    & $0.003$ & (**)           \\
 && \textit{Level Shift}   & $11.97$& $2.29$    & $p<0.001$ & (***)           \\
\hline

\multirow{5}{*}{\textit{\textbf{31}}}& \multirow{5}{*}{ARIMA(1,2,2)} & \textit{Constant}   & $-5.38$    & $2.66$      & $0.046$ & (*)    \\
&& \textit{MA (1)}   & $0.64$& $0.09$    & $p<0.001$   &  (***)         \\
&& \textit{MA (12)}   & $0.85$ & $0.11$    & $p<0.001$ & (***)            \\
&& \textit{Ramp}   & $-123.07$& $32.69$    & $0.003$ & (***)           \\
&& \textit{Level Shift}   & $21.57$& $6.51$    & $0.002$ & (**)           \\
\hline

\multirow{3}{*}{\textit{\textbf{03}}}&\multirow{3}{*}{ARIMA(1,2,2)} & \textit{MA (1)}   & $0.89$    & $0.05$      & $p<0.001$ & (***)    \\
 && \textit{MA (12)}   & $-0.65$ & $0.10$    & $p<0.001$ & (***)            \\
 && \textit{Level Shift}   & $-3.71$& $1.16$    & $0.002$ & (**)           \\
\hline

\multirow{2}{*}{\textit{\textbf{30}}}&\multirow{2}{*}{ARIMA(1,2,1)} & \textit{MA (12)}   & $0.69$    & $0.09$      & $p<0.001$ & (***)    \\
 && \textit{Level Shift}   & $-63.04$ & $6.54$    & $p<0.001$ & (***)            \\
\hline

\multirow{2}{*}{\textit{\textbf{13}}}&\multirow{2}{*}{ARIMA(1,1,1)} & \textit{MA (1)}   & $0.91$    & $0.04$      & $p<0.001$  & (***)       \\
&& \textit{Level Shift}   & $-5.67$& $1.16$    & $p<0.001$ & (***)           \\
\hline

\multirow{3}{*}{\textit{\textbf{23}}}&\multirow{3}{*}{ARIMA(2,1,0)} & \textit{AR (1)}   & $0.46$    & $0.10$      & $p<0.001$ & (***)    \\
 && \textit{AR (12)}   & $-0.64$ & $0.11$    & $p<0.001$ & (***)            \\
 && \textit{Level Shift}   & $-41.73$& $19.19$    & $0.032$ & (*)           \\
\hline

\end{tabular}
\end{adjustbox}
    \caption{ARIMAX models for evaluating the policy impact on opioid drugs for classes with significant intervention coefficients.}
    \label{tab:Overall Policy Effect}
\end{table}

\newpage
\section{Discussion}
The current study finds that the opioid prescription limit legislation in South Carolina resulted in a statistically-significant decrease in overall opioid prescriptions. However, the impact is not the same on each group classified according to the relative locations of patients, prescribers, and the dispensers. For example, the average daily MME for the group of isolated prescribers and patients traveling more than 500 miles, increased from 46.69 (95\% CI [46.35, 47.03]) to 48.49 (95\% CI [47.31, 49.67]) (See Table~\ref{tab:MMEperDay_PrePostPolicy}). This raises the question of whether patients are traveling longer distances to obtain prescriptions to avoid the legislative limits. 

The rationale behind opioid-limiting legislation is that decreasing exposure to opioids among uninformed patients, as well as decreasing the supply of available opioids in the community for misuse, may aid in curtailing the opioid epidemic \citep{sedney2021assessing, maclean2020economic, davis2019laws}. However, there could be unintended consequences \citep{maierhofer2022intended}, such as motivating those with opioid use disorders to access the illicit drug market \citep{lee_systematic_2021}, or obtaining opioid drugs from friends \citep{bicket2017prescription} potentially increasing overdose mortality.

The findings of this study indicate that the groups within the classification system exhibit different patterns of opioid consumption. Specifically, it was observed that groups in which patients travel distances exceeding 500 miles to obtain drugs from distant pharmacies have a consumption level higher than 90 MME per day, and are distinct from other groups. Previous research has established that patients engaging in illegal opioid activities such as doctor shopping, which involves visiting multiple healthcare providers to acquire medication illegally \citep{sansone_doctor_2012}, are more likely to travel long distances and across state borders \citep{cepeda_distance_2013}. Our classification system flags such transactions and may help identify illegitimate activities. We think this would be an interesting follow-up study to our work.

Furthermore, the study shows that the prescription limit policy helped reduce overall opioid prescription volume, as evidenced by the decrease in average daily MME from 53.68 (95\% CI [53.33,54.02]) to 51.09 (95\% CI [50.74,51.44]). The ARIMA and ARIMAX model results reveal that the pre-policy prescription data is not accurate in forecasting the post-policy trend, and the intervention has a significant impact when included in the model. Furthermore, the examination of the control group--benzodiazepine prescriptions--indicate that the policy had minimal effect on these drugs, providing further evidence that the policy had the desired impact on the targeted group of drugs.

The research demonstrates that the policy has been effective for a majority of transactions where the total distance traveled is less than 250 miles. Specifically, it is found that there is a difference between pre-intervention model predictions and post-intervention actual values for all transactions in this category, but a significant impact of the intervention determined by the use of ARIMAX models was only identified in one class (class 03) within this group, which also had the highest frequency in the data. This implies that the policy is more successful in transactions that take place within the state borders, as it is more likely to achieve its intended impact on the target population.

The study yielded an interesting finding related to prescriber isolated classes, specifically a discrepancy between the prescription volume predicted values from ARIMA models and actual values in classes 01, 11, and 31. Interestingly, the mismatch is desired in class 01 but undesired in groups 11 and 31, where predicted values are below actual values and actual values experience an increase after the policy's adoption. Additionally, ARIMAX models indicate that groups 11 and 21, in which patients visit far-away doctors to obtain their drugs, display unusual behavior and the policy's effect is significant in the opposite direction. This suggests that after the policy's adoption, patients in South Carolina visit doctors in states with looser regulations to obtain their drugs. These states may not necessarily be neighboring states, as many neighboring states implemented similar laws during the 2017-18 period. This study did not take into account the policy adoption time in other states, but this could be addressed in future work. Overall, we caution policy makers against undesired results that are not visible when analyzing aggregated data but can be observed only with proper classification. It's also worth mentioning that classes 11 and 31 only make up 2.51\% of the data records, and the undesired impact of legislation in these groups may not be visible when aggregated with other classes.

The study found that the policy lower impact on dispenser isolated groups, where patients visit a local doctor but obtain their prescription from a distant pharmacy. This is significant as these groups have a much higher daily MME, which is considered highly hazardous by the CDC (specifically groups 22 and 32 with an average consumption of over 90 MME per day). Therefore, it is important to design a policy to reduce prescription volume in these classes. The reason for the unusual prescription volume among these groups is not known and is beyond the scope of this research, but it is clear that the specific prescription limit policy implemented in South Carolina is not able to prevent distant pharmacies from dispensing high volumes of prescribed opioids to patients in South Carolina. It is worth noting that this is not a drawback of the policy under study as it was not specifically designed to address these groups.

Overall, the study finds that policy effectiveness depends on many factors and that the total distance and relative location of the drug supply stakeholders are among the important ones. Policy makers should not overlook the unintended consequences of legislation and evaluate the policy's effectiveness across different population groups.

\section{Conclusion} 
The current study a statistically-significant found evidence of a 4.82\% reduction in overall opioid prescription volumes, as measured by daily MME, over a 43-month period following the implementation of the NarxCare prescription limitation law in South Carolina. Furthermore, there was no significant change in benzodiazepine prescription volume--the control group which was not targeted by legislation. Analysis of the data revealed a decrease in transactions that occurred within a 250-mile radius, likely within the state's borders, which constitute over 90\% of total transactions in the state. However, transactions involving patients who received prescriptions from far-away doctors increased following the law's passage. This suggests a potential drawback of the law that is not immediately apparent without a proper classification of the data. Additionally, the law did not significantly impact prescriptions filled by distant pharmacies, and this group has the highest consumption level among all. We recommend greater oversight of out-of-state doctors or a nationwide regulation of opioid drugs to address illicit purchases. It is important to note that the effect of this legislation on diverted or misused opioids was not evaluated in this study.

\bibliographystyle{elsarticle-harv}\biboptions{authoryear}
\bibliography{cas-refs}

\end{document}